\documentclass{mn2e}  
\usepackage{amssymb} 
\input epsf.sty  
\newif\ifAMStwofonts  
\AMStwofontstrue  
  
\def\Rs{$R_{\rm Schw}$}

\def\Agata{R\' o\. za\' nska~}
  
\begin{document}  
  
\title{Irradiation of accretion discs in active galactic nuclei due to warm 
absorber}  
\author[Loska et al.]  
{Z. Loska$^1$,  B. Czerny$^1$, R. Szczerba$^2$\\  
 $^1$N. Copernicus Astronomical Center, Bartycka 18, 00-716 Warsaw, Poland\\ 
 $^2$N. Copernicus Astronomical Center, Rabia\' nska 8, 87-100 Toru\' n, 
 Poland } 
\maketitle  
\begin{abstract}  
The presence of the warm absorber of considerable optical depth is seen 
in many AGN. We show that this medium may affect significantly the 
optical/UV spectrum of an AGN by backscattering a fraction of the 
total radiation flux towards the disc surface. We consider in detail the
case when the disk extends down to a marginally stable orbit, 
all the emission comes from the disk surface and the scattering medium
forms a cone around the symmetry axis.
Disc irradiation results in 
much flatter optical/UV continuum than predicted by standard disc models.
The effect depends both on the total optical depth of the warm absorber and 
on the specific density distribution of this medium so the analysis of the 
optical/UV continuum allows to obtain  constraints for the warm absorber
complementary to those obtained from the soft X-ray data analysis.
We give results for two exemplary sources - RE J1034+396 and PG1211+143, and
for the bluest composite quasar spectrum of Richards et al. obtained from SDSS.
  
\end{abstract}  
  
  
\section{Introduction}
\label{intro} 

Optical/UV spectra of bright active galactic nuclei (AGN) are usually
interpreted as originating from an accretion disc surrounding the
central massive black hole (Lynden-Bell 1969, Shields 1978, Malkan \& 
Sargent 1982; for a review, 
see e.g. Koratkar \& Blaes 1999,
Czerny 2003). However, both the shapes of the radiation
spectra  and the
variability studies (e.g. Wanders et al. 1997, Collier et al. 
1998 for NGC 7469; see Ulrich, Maraschi \& Urry 1997 for a review) 
suggest that significant fraction of radiation emitted
at a given radius is due to the reprocessing of the radiation 
dissipated elsewhere. 

Most of the radiation is produced in the deep potential well, close to the
black hole. In quasars and Seyfert 1 galaxies part of this radiation 
reaches an observer directly. However, part of the radiation may be
recaptured by the outer disc region.

The irradiation of the disc may be direct or indirect. In the first case
the disc must be exposed to the radiation flux generated in the
inner region. In the second case there may be a scattering medium which
redirects a part of the radiation back towards the disc.

The direct irradiation of AGN accretion discs was discussed by several
authors (e.g. Krolik et al. 1991, Courvoisier \& Clavel 1991, Collin-Souffrin 
1991, Rokaki et al. 1993, Loska \& Czerny 1997, Kurpiewski et al. 1997,
Sincell \& Krolik 1997, Collin \& Hur\' e 1999, Nayakshin et al. 2000, 
Ballantyne et al. 2001,
Nayakshin \& Kazanas 2002, Soria \& Puchnarewicz 2002, \Agata et al. 2002,
Chiang \& Blaes 2001, 2003, Chiang 2002). 
Most of these papers were actually devoted to the irradiation of the inner 
disc by the hard X-rays generated in a lamp-post or disc corona.
Much of the research had been 
done in the context of X-ray binaries (e.g. Raymond 1993, Dubus et al. 2001,
Jimenez-Garate et al. 2002) 
or cataclysmic
variables (Meyer \& Meyer-Hofmeister 1982, Smak 1989) 
and the methods are directly
applicable to AGN discs but the results may not apply quantitatively
since the disc shape depends on the mass of the central object, as already
seen from the classical paper of Shakura \& Sunyaev (1973).

The indirect irradiation was also considered in several papers (Ostriker 
et al. 1991, Murray et al. 1994, Kurpiewski et al. 1997; also Esin et al. 1997
in the case of galactic sources). These authors analyzed the scattering
of the radiation in the accretion disc corona. 

In the present paper we consider the {\it scattering} of the disc radiation
by the warm absorber. The warm
absorber medium, seen in many AGN (for a review, see e.g. Kriss 2004), 
is a highly ionized plasma so it
predominantly scatters the passing radiation, with absorption important
only in the soft X-ray band. The optical depth of the warm absorber
for scattering is difficult to measure but estimated values can be as high
as 0.33 or more for PG1211+143 (Pounds et al. 2003a),  0.26 or more for
PG PG 0844+349(Pounds et al. 2003b), 0.20 or more for PG 1402+261 (Reeves,
Porquet \& Turner 2004). 
Arav et al. (2003) show by analyzing OVI UV absorption and OVI
X-ray absorption in NGC 5548 that simple modelling gives discrepant results 
and in general heavily underestimates the amount of outflowing material.

 Therefore, a significant
fraction of the radiation
flux emitted by the central region can be in such way redirected towards the
outer parts of the accretion disc and thermalized there. Therefore, the 
presence of highly ionized warm absorber of considerable optical depth must
modify the IR/optical/UV disc spectra. We can apply this effect either to 
explain the departure of the observed spectrum from the simple prediction of 
the accretion disc model or, if no such clear departure is seen, as an
independent way to put limits to the optical depth and the covering factor of 
the warm absorber.   

The effect should be present in all AGN, but it is potentially most important
when the direct disc irradiation by an X-ray source is relatively weak.
We assume the accretion flow
proceeds through a standard disc which extends down to the marginally stable 
orbit. We neglect any hard X-ray emission.
We calculate the 
optical/UV spectra of the accretion discs taking into account the scattering
of the disc flux by the warm absorber as well as the effect of the direct
self-irradiation of the disc. 

As an application, we show the spectra models for two NLS1: RE~J1034+396 and 
PG~1211+143, and for the composite 1 quasar spectrum of Richards et al. (2003).

\section{Method}

We consider the self-irradiation of the disc both due to direct 
self-illumination and due to radiation scattering by the warm absorber.

In the case of an extended scattering medium,
there is practically no dependence of the result on the disc shape. However,
the direct irradiation is very sensitive to the adopted disc shape: 
the disc must 
flaring in order to have its surface exposed to the central source. 
We must specify the disc model appropriate for an AGN in order to see which 
disc parts actually are irradiated. 

Therefore, in Sect.~\ref{sect:shape1} we describe the disc model, 
in Sect.~\ref{sect:direct1} we show the formulae for the direct 
irradiation, in 
Sect.~\ref{sect:scatt1} we
describe the warm absorber model and the method of computing the incident
flux at a given disc radius, and in Sect.~\ref{sect:spec} we describe the
method of calculating disc spectrum.  

\subsection{Disc shape}
\label{sect:shape1}

We use the stationary Keplerian disc model of \Agata et al. (1999). The disc
thickness is calculated by solving the equations of the disc vertical 
structure. Viscosity is described assuming the standard $\alpha$ approach
of Shakura \& Sunyaev (1973), i.e. we assume that the viscous torque is 
proportional to the total (gas + radiation) pressure. Recent studies of 
magneto-rotational instability generally supports the validity of such scalling
(e.g. Turner 2004). Estimates based on the MHD simulations of the MRI
instability indicate the average value of the effective viscosity parameter 
$\alpha$ about 0.02 (Winters et al. 2003), and similar values of 
$\alpha$ between 0.01 and 
0.03 were derived from the analysis of the AGN variability 
(Starling et al. 2004).

The disc vertical structure (as a function of the distance $z$ from 
the disc equatorial plane) is calculated by
solving the equations of viscous energy dissipation, hydrostatic
equilibrium, and energy transfer:
\begin{equation}
{d F \over d z} = \alpha P_{\rm gas} (-{d\Omega \over d r})
\end{equation}
\begin{equation}
{1 \over \rho} {d P \over d z} = -\Omega^{2} z
\label{eq:hydro}
\end{equation}
\begin{equation}
{d T \over d z} = -{3 \kappa \rho \over 4 a c T^{3}} F_{\rm l}.
\end{equation}
Here $\Omega$ is the Keplerian angular velocity, $a$ and $c$ are physical 
constants.
$F_{\rm l}$  
is the energy flux transported locally in the direction perpendicular 
to the equatorial plane and carried either by radiation or by convection:
\begin{eqnarray}
F_{\rm l} = & F_{\rm rad} ~~~~~~~ & \nabla_{\rm rad} \le \nabla_{\rm ad}\nonumber \\ 
F_{\rm l}  = &  F_{\rm rad} +F_{\rm conv} ~~~~~ & \nabla_{\rm rad} > \nabla_{\rm ad}
\end{eqnarray}
These equations, supplemented  by the equation of state, describe the 
temperature, density and pressure profiles, $T(z)$, $\rho(z)$ and $P(z)$.

The frequency-averaged opacity $\kappa$ is taken to
be the Rosseland mean and includes the electron scattering, free-free
and bound-free transitions. The opacity tables are from Alexander,
Johnson \& Rypma (1983) and Seaton et al. (1994).  The presence of
dust and molecules is included in the opacity description.  The
details of the model were discussed in Pojma\'nski 1986 and 
R\'o\.za\'nska et al. (1999).

Disc thickness is determined iteratively, from the condition that the energy
flux dissipated inside the disc is equal to the energy flux given by assumed
accretion rate through the standard formula
\begin{equation}
F(H,r) = {3 G M \dot M \over 8 \pi r^3}(1 - \sqrt{r \over 3 R_{Schw}})
+ F_{inc}(H,r)(1 - A).
\label{eq:diss}
\end{equation}
This means that we use the Newtonian approximation for a non-rotating black 
hole. Consequently, we use the dimensionless accretion rate in
Eddington units, 
$\dot m = \dot M / \dot M_{\rm Edd}$, with the
efficiency of 1/12:
\begin{equation}
\dot M_{\rm Edd} = 2.66  { M \over 10^8 M_{\odot}} [M_{\odot}/yr].
\end{equation}

The effect of irradiation is included only through the additional term 
present in Eq.~\ref{eq:diss} (see  e.g. Tuchman et al. 1990), 
where $F_{inc}$ is either $F_{direct}$ or
$F_{sc}$, or both, since we 
concentrate on the overall spectral shape, particularly in the optical/UV 
band. 
This term describes the absorbed fraction of the
external incident flux. The method of determination of this flux is 
described in the Sections~\ref{sect:direct1} and \ref{sect:scatt1}. The albedo,
$A$, is assumed to be independent from the photon energy and its 
energy-independent mean value is assumed to be equal to 0.2, appropriate
for cold disc surface (see e.g.  Haardt \& Maraschi 1991).

\subsection{Direct irradiation}
\label{sect:direct1}

Computing the direct irradiation we follow the approach of Fukue (1992).

We calculate the irradiation flux from the given distribution of the emitted
flux, $F_{em}(r)$ and the disc shape, $H(r)$.
The incident flux at a given disc radius, $r_o$, has the form of the integral 
over the disc surface
\begin{eqnarray}
F_{direct}(r_o)  = {4  \over \sqrt{1 + ({dH_o \over dr})^2}}  \nonumber \\
 \times \int_0^{\pi}
\int_{R_{in}}^{R_{out}} {F(H,r) \cos i \cos i_o 
\sqrt{1 + ({dH_o \over dr})^2} r dr d \phi \over D_s^2},
\end{eqnarray}
with an additional condition that the contribution to the integral is included
only if 
\begin{equation}
 \cos i > 0, ~~~~~~\cos i_o > 0.
\end{equation}

\begin{figure}  
\epsfxsize = 90 mm  
\epsfbox[20 500 570 750]{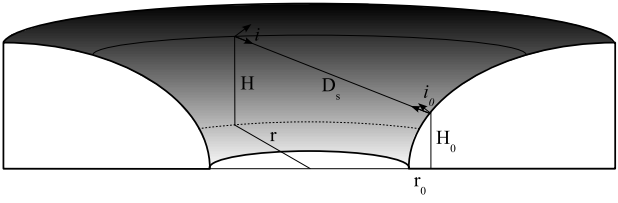}  
\caption{The schematic picture of the self-irradiation of the accretion 
disc. 
\label{fig:draw1}}  
\end{figure}

The incident flux 
$F_{inc}=F_{direct}$ 
is calculated at the radius $r_o$, $r$ is the current radius of
integration over the disc surface, $H_o$ is the disc thickness at $r_o$ while
$H$ is the disc thickness at the radius $r$, $i_o$ and $i$ are 
the corresponding angles between
the normal to the disc surface and the line joining two disc surface points,
as marked schematically in Fig.~\ref{fig:draw1}. The distance $D_s$ between the
irradiating and irradiated surface element is given by
\begin{equation}
D_s^2 = r^2 + r_o^2 + (H - H_o)^2 - 2 r r_o \cos \phi,
\end{equation}
the angle $i$ is determined by
\begin{equation}
\cos i = { {d H \over dr} (r - r_o \cos \phi) + H_o - H \over D_s \sqrt{1 + {({dH \over dr})^2}}},
\end{equation}
and the angle $i_o$ is determined by 
\begin{equation}
\cos i_o = { {d H_o \over dr} (r_o - r \cos \phi) + H - H_o \over D_s \sqrt{1 + {({dH_o \over dr})^2}}},
\end{equation}.

Radiation flux from the disc is determined by the Eq.~\ref{eq:diss}.
Since the effect of irradiation depends on the shape of the disc, and both
the disc shape and the local emissivity depend on irradiation, the result is 
obtained through consecutive iterations, starting from unilluminated disc. 
Usually 2 - 3 iterations are enough to
provide the result with the accuracy better than 1\% .

\subsection{Scattering by the warm absorber}
\label{sect:scatt1}

The idea of redirecting some of the disc radiation flux emitted in the
central region towards the outer part through scattering by the warm absorber
was suggested by Czerny et al. (2003). It could be seen from Fig.~3 of this
paper that the warm absorber absorbs efficiently only in the soft X-ray band
while for lower frequency photons the opacity is equal to that of electron
scattering and this effect dominates the overall extinction in the optical/UV 
band. The key parameter is
the ionization parameter $\xi$ defined as
\begin{equation}
\xi = {L \over nR^2},
\end{equation} 
where $L$ is the luminosity of the central source, $n$ is the number density
of the shell and $R$ is the inner radius of the shell. 

We checked the effect using the code {\sc cloudy}, version 94.0, for a 
spherical shell geometry.
The reflected continuum is redirected towards the disk and the transmitted
continuum is seen by an observer. 
For $\xi=10^4$ the warm absorber imprints significant features in
the X-ray band
into the transmitted spectrum which allow to determine the hydrogen column
of the absorbing medium. Narrow absorption lines form also in the UV 
band. At the same time the overall reflected spectrum still
is not much different from the intrinsic continuum. In the case of almost
fully ionized absorber ($\xi=10^5$) electron scattering dominates so 
strongly that actually almost no spectral features are seen. It
supports even better our later approach based on the negligence of absorption
processes. At the same time, it  illustrates potential problems in detecting
warm absorbers. An extremely strongly ionized medium with large optical depth
may well exist in an object and still escape detection by analysis of the
soft X-ray data.


The presence of the dust within the warm absorber complicates the issue. 
As an example, we 
considered a case with $\xi = 10^4$, but we allowed for
an additional fraction of dust, at amount of 1\% of the standard
gas to dust ratio for the interstellar medium in our Galaxy, and of the same
composition. Electron scattering is still quite effective, and the dust
also partially scatters the incident photons. The net effect is strong
reddening in the transmitted continuum and relatively minor suppression
of the UV part in the reflected spectrum. Details of those effects depend
significantly on the dust composition. The possible presence of the dust in the
warm absorber is discussed in a number of papers (e.g. Komossa \&
Breitschwerdt 2000, Mason et al. 2003,
Ballantyne, Weingartner \& Murray 2003). 

In the present paper we concentrate on objects which are not heavily
reddened in UV and we consider the warm absorber 
medium as purely scattering fully ionized plasma.


The origin, the geometry and the physical state
of the warm absorber are still under discussion. As for the source of material,
ideas include accretion disk winds/Broad Line Region clouds (Elvis 2000, 
Kriss et al. 2003, Blustin et al. 2003), inner edge of the
dusty/molecular torus (Krolik \& Kriss 2001) and radiation driven clouds
constituting the Narrow Line Region (Crenshaw et al. 2000). 
King \& Pounds (2003) go as
far as suggesting that actually the outflow from the direct vicinity of an
accreting black hole may be  strong enough to form an optically thick shell 
that Comptonize the emission from the 
directly unseen inner source.

Narrow absorption lines in 
many spectra
of Seyfert 1 galaxies indicated a clumpy medium with numerous colder clumps
embedded in an fully ionized hotter medium (Crenshaw et al. 1999; 
Kriss 2002; Krolik \& Kriss 2001, Blustin et al. 2003) 
while 280 ksec XMM-Newton 
spectroscopy of NGC 3783 (Behar et al. 2003) suggested mostly uniform medium, 
at distances up to 2.8 pc from the nucleus.

The optical depth of the warm absorber is poorly constrained since the 
directly measured column density of specific ions must be supplemented with
modelling of the ionization state of the plasma in order to obtain the
hydrogen column. 

If cooler clumps are embedded in a fully ionized medium the
contribution of this medium is particularly difficult to estimate while it may
significantly enhance the total optical depth of the medium for electron 
scattering.

Detection of the variability in the OVIII hydrogen column of MCG-6-30-15 (Otani
et al. 1996) allowed to estimate the distance of the warm absorber as smaller
than $10^{17}$ cm but the lack of variability in the OVII edge suggested that 
the warm absorber is rather spatially extended, with highly ionized flow 
closer in and less ionized flow further out, as visualized by Blustin et al.
(2003). Models based on evaporation of the inner torus suggest distances of
a fraction of a parsec while models with disk outflow suggest distances of
a few Schwarzschild radii for fully ionized flow.

Radial velocities of warm absorber features of 
order of a few hundreds km/s determined in a number of sources, 
if interpreted as
roughly Keplerian speed, may indicate rather large distances and the connection
with dusty torus (e.g. Ashton et al. 2004). Large velocities claimed to be
seen in PG1211+143 (Pounds et al. 2003a) and PG0844+349 (Pounds et al. 2003b) 
suggest much smaller values, although
identification of the absorption lines and consequently, determination of the 
outflow velocity, may be questioned (Kaspi 2004).

\begin{figure}
\epsfxsize=9.0cm \epsfbox[50 400 500 830]{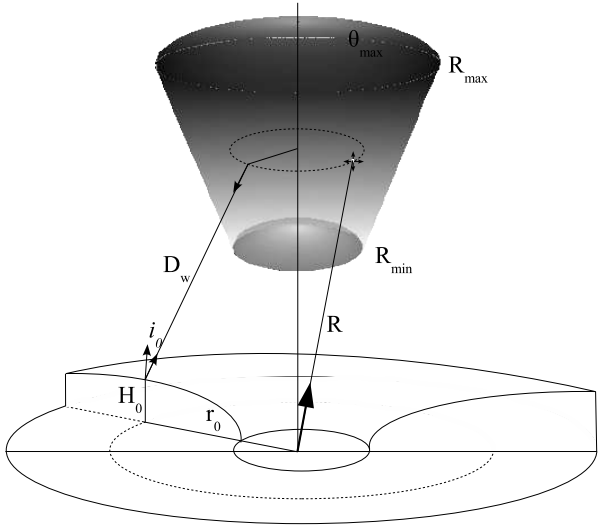}
\caption{The schematic picture of scattering of the disc emission by the 
warm absorber.}
\label{fig:draw2}
\end{figure}

We assume that the warm absorber has an axially-symmetric conical shape. It
extends from the inner radius, $R_{min}$, to the outer radius, $R_{max}$, with
the opening angle of the cone given by $\theta_{max}$. The schematic picture is shown
in Fig.~\ref{fig:draw2}.

The density of the warm absorber depends on the radial 
distance from the center, $R$ in the form
\begin{equation}
 n(R) = n_o ({R \over R_{min}})^{-\beta},
\end{equation}
where the density at the inner edge of the warm absorber is conveniently 
expressed as
\begin{equation}
n_o = {\tau_{tot} \over \sigma_T \int_{R_{min}}^{R_{max}} ({R \over R_{min}})^{-\beta}dR},
\end{equation}
since the total optical depth of the warm absorber, $\tau_{tot}$, is relatively easily 
estimated from observations. We can also include in our scheme the dependence 
of the density distribution on $\theta$. 

Since the warm absorber is an extended medium, we neglect the extension of the
inner parts of the disc where most of the energy is generated and we assume 
that the whole radiation comes from a point-like source. Its luminosity, $L$, 
is determined by the accretion rate
\begin{equation}
L = \dot m L_{Edd}.
\end{equation}

We take into account the finite optical depth of the warm absorber so the local
emissivity of this medium due to the electron scattering is given by
\begin{equation}
j(R,\theta) = \alpha_T(R) { 2 L \cos \theta\over4 \pi R^2 \exp [{- \int_{R_{min}}^R {\alpha_T(s)ds]}}},
\end{equation}
where the opacity $\alpha_T$, is the Thompson opacity 
($\alpha_T(R) = \sigma_T n(R)$).

This extended medium returns the fraction of the scattered radiation towards 
the disc. The resulting illumination is calculated as
\begin{eqnarray}
F_{sc}(r_o) = \nonumber \\
{1 \over 4 \pi} \int_0^{2 \pi} \int_{R_{min}}^{R_{max}} \int_0^{\theta_{max}} { j(R,\theta) R^2 \sin \theta \cos i_o \over D_w^2} d\theta dR d\phi,
\end{eqnarray}
where 
\begin{equation}
\cos i_o = { {d H_o \over dr} (r_o - R \sin \theta \cos \phi) + R\cos \theta - H_o \over D_w \sqrt{1 + {({dH_o \over dr})^2}}},
\end{equation}
and 
\begin{equation}
D_w^2 = r_o^2 + R^2 + H_o^2 - 2 R (r_o \sin \theta \cos \phi + H_o \cos \theta). 
\end{equation}

\subsection{Accretion disc spectrum}
\label{sect:spec}

Accretion discs in high Eddington rate 
AGN are optically thick and they thermalize the absorbed
radiation efficiently. The models of the emitted local spectrum, however,  
show some departure from the local black body approximation due to the role 
of the Compton scattering in the disc outer layers (e.g. Shimura \& Takahara
1995, Madej \& \Agata 2004). This departure is usually well described
in terms of the colour temperature to effective temperature ratio. In various
models this ratio varies from 1.7 - 2.0 (Shimura \& Takahara 1995, Ross \&
Fabian 1996) to 1.8 - 2.6 
(Merloni, Fabian \& Ross 2000).
Madej \& \Agata (2004) find $T_{col}/T_{eff} < 2 $ in their set of neutron
star atmosphere models.
Recent observational studies of galactic sources in their high states 
(Gierli\' nski \& Done 2004a) suggest
that the value of $\sim 1.8$ indeed well represents the effect. 
We adopt this value
in our description of the inner part of the disc where electron scattering
dominates the opacity. In the outermost region the disc is not so strongly
ionized and the emission is expected to be close to a black body emission. 
Therefore, we describe the local spectrum as 
\begin{equation}
F_{\nu} = {cos i \over 4 \pi D^2}\int_{r_{in}}^{\infty} 
\pi B_{\nu}(T_{col}) 2 \pi r dr,
\end{equation}
where the effective temperature is calculated from the total (dissipated +
thermalized) flux
\begin{equation}
\sigma T_{eff}^4 = F(H,r),
\end{equation}
and to ensure the continuity between the inner and the outer disc region, 
we assume
\begin{equation}
T_{col}/T_{eff} = 2.228 - 0.041 (r/R_{Schw})~~{\rm if}~ r < 30 R_{Schw}
\end{equation}
and $T_{col}= T_{eff}$ at larger radii.

\section{Results}

\subsection{Disc shape}
\label{sect:shape2}

\begin{figure}  
\epsfxsize = 90 mm  
\epsfbox{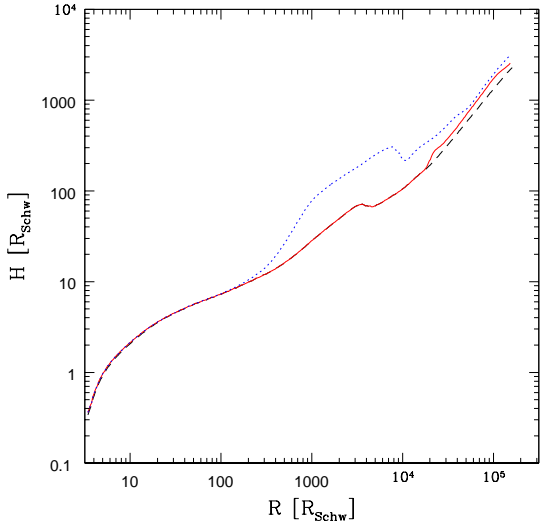}  
\caption{The dependence of the disc thickness on the disc radius without
the effect of self-irradiation (dashed line), with the self-irradiation
(continuous line) and with the back-scattering flux by the warm absorber
(dotted line). Disc
parameters: $M=10^6 M_{\odot}$, $\dot m = 0.5$, $\alpha = 0.02$, warm 
absorber parameters: $\tau_{tot} = 0.6$, $\beta = 0$, $R_{min} = 
1000 R_{Schw}$, 
$R_{max} = 1200 R_{Schw}$, $\theta_{max} = \pi/3$. 
\label{fig:grubosc}}  
\end{figure}  
  
An exemplary shape of AGN accretion disc is shown in Fig.~\ref{fig:grubosc}. 
It is much more complex than the simple prediction of Shakura \& Sunyaev (1973)
for radiation-pressure dominated region. The departure is not important from
the point of view of irradiation by the extended warm absorber but 
it is important for 
self-irradiation. Even in the innermost radiation pressure dominated 
region the
disc thickness is not constant but rises slowly. Further out, 
it rises faster as
the radiation pressure drops. Some wiggling is also visible - this 
is due to the
complex opacities used for the description of the vertical structure.
The opacity of the outermost weakly ionized disc region is dominated by 
molecules and absorption by dust, so the disc is never optically thin.
 
\begin{figure*}
\parbox{\textwidth}{
\parbox{0.5\textwidth}{
\epsfxsize=0.45\textwidth
\epsfbox[18 200 600 720]{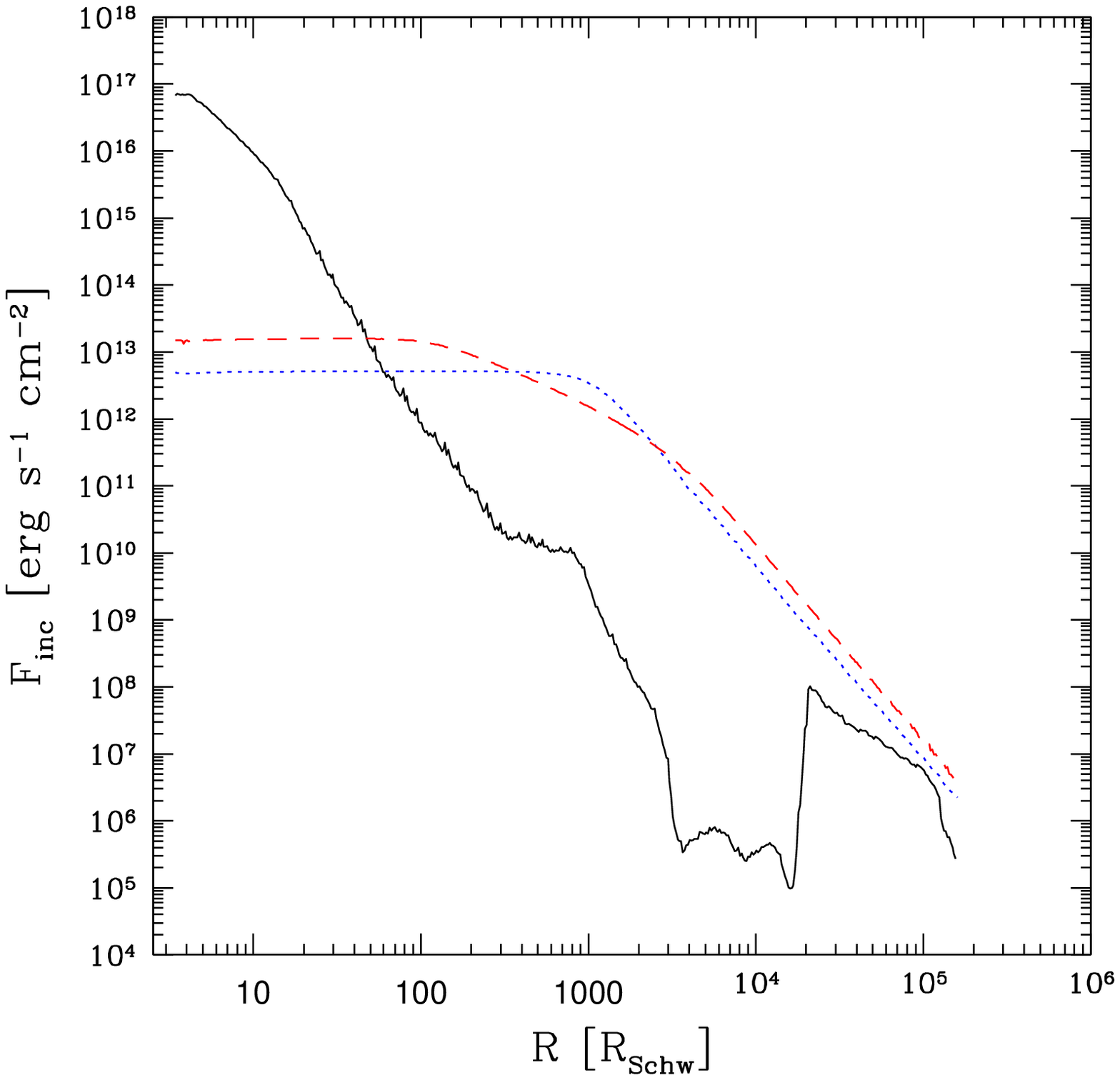}}
\hfil
\parbox{0.5\textwidth}{
\epsfxsize=0.45\textwidth 
\epsfbox[18 200 600 720]{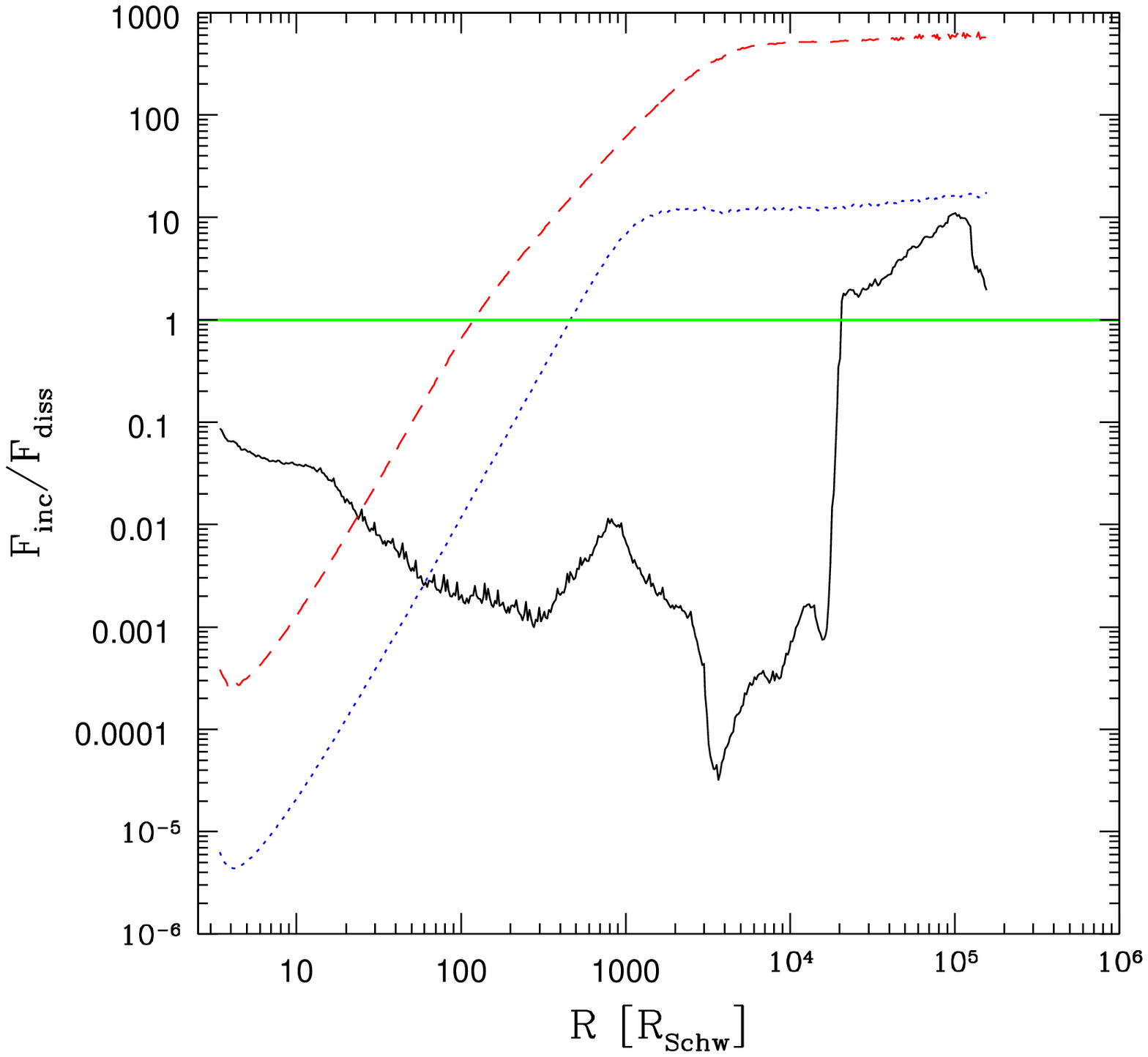}}
}
\caption{The irradiation flux (left panel) and the ratio of the irradiation 
flux to the locally dissipated flux for the same model as in 
Fig.~\ref{fig:grubosc}. Continuous line - self-irradiation; 
dotted line - back-scattering by the warm absorber. We also show the 
back-scattered flux (dashed line) obtained for the warm absorber 
extending from $R_{min} = 100 R_{Schw}$ to $R_{max}=5000 R_{Schw}$, and with
the same optical depth $\tau_{tot}= 0.6$ as before. Thick horizontal line marks
the region where irradiation dominates.}
\label{fig:flux}
\end{figure*}

In particular,
the drop in the disc height at $\sim 3000 R_{Schw}$ happens in the region of 
ionization instability (Lin \& Shields 1986). 
Here we neglect the issue of the 
time-dependent 
disc behaviour. It is most probably justified for high accretion rate 
objects since
no evaporation of an inner disc is expected and ionization instability 
reduces to small
flickering for a constant viscosity law (Siemiginowska, Czerny \& Kostyunin 
1996,  Janiuk et al. 2004). 
If the irradiation by the warm absorber is included, this region 
is shifted outward (see Fig.~\ref{fig:grubosc}, dotted line). The disc
thickness is increased by a factor of 3 at distances $\sim 1000 R_{Schw}$
from the black hole since irradiation was most efficient there for an
adopted model of the warm absorber.  
Self-irradiation (see Fig.~\ref{fig:grubosc}) has
a very small impact on the disc thickness, visible only at very 
large distances, above $10^4 R_{Schw}$.

If accretion rate is significantly lower than used in Fig.~\ref{fig:grubosc},
i.e. $\dot m = 0.05$ instead of $\dot m = 0.5$ then the disc thickness in the
innermost (radiation pressure dominated) part is roughly by a factor 10
smaller, the wiggle in the disc shape (i.e. partial ionization instability
region) moves in roughly by a factor of 3, and the disc thickness 
in the outer region is almost unchanged. The effect of the self-irradiation
and irradiation due to the scattering by the warm absorber modify the disc
thickness in roughly the same way as before.  

The change in the properties of the warm absorber is reflected in the
disc thickness. Lower/higher total opacity leads to smaller/larger increase
in the disc height at distances where the irradiation is important. 

The density profile of the warm absorber specifies at which distance the 
irradiation is most effective ($R_{min}$, $R_{max}$ or a broad range between the
two). However, performing subsequent iterations between irradiation and the 
disc thickness affect the final irradiation flux rather weakly.

\subsection{Radial dependence of the irradiating flux}

\subsubsection{Direct irradiation}

An example of the incident radiation flux in the absence of the warm absorber 
is shown in Fig.~\ref{fig:flux} (continuous line). Model parameters are the 
same as used in Fig.~\ref{fig:grubosc}. 

The flux is large at small radii, but it is nevertheless small in
comparison with the locally dissipated flux in this region. Self-irradiation 
does not create energy, it only effectively modifies slightly the angular
dependence of the radiation leaving the central region. However, as it
is well known, irradiation becomes important at large distances 
(e.g. Hoshi \& Inoue 1988). The sudden jump in the flux is again a 
consequence of the rise
of the disc thickness at the end of the partial ionization zone (see
Fig.~\ref{fig:grubosc}).

\begin{figure}  
\epsfxsize = 90 mm  
\epsfbox{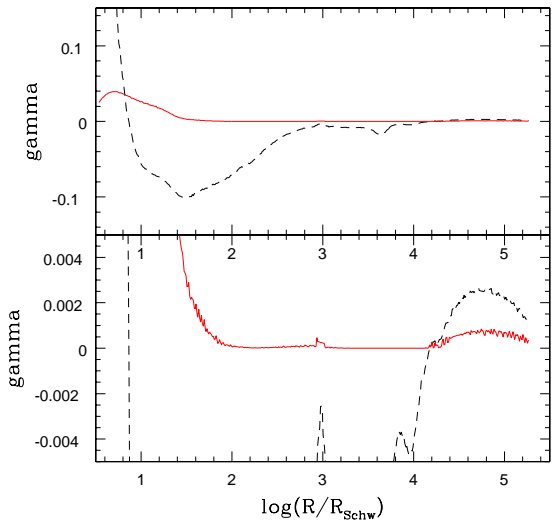}  
\caption{The dependence of the $\gamma$ factor on the disc radius. Continuous 
line: illumination by extended disc. Dashed line: point-like source 
approximation given by Eq.~\ref{eq:gamma2} for n=1. Long dashed line: point-like 
source approximation given by Eq.~\ref{eq:gamma2} for n=2. 
Model parameters as in Fig.~\ref{fig:grubosc}. Negative
$\gamma$ means that a point-like source located at the center would not 
irradiate the disc surface. 
\label{fig:gam_factor}}  
\end{figure}  

Sometimes, for simplicity, the irradiation is calculated assuming the
point-like source instead of an extended inner disc. Irradiation flux is in
this approximation given by (e.g. King 1998, King \& Ritter 1998)
\begin{equation}
F_{inc} = {L \over 4 \pi r^2} \gamma,
\label{eq:gamma1}
\end{equation}
where 
\begin{equation}
\gamma = ({h \over r})^n({d \ln h \over d \ln r} - 1).
\label{eq:gamma2}
\end{equation}

The form of Eq.~\ref{eq:gamma1} is general so it is illustrative to express
the distribution of the irradiation flux determined from our model also
as a radial dependence of the dimensionless quantity gamma. The value $n = 1$
is expected to be better for neutron star systems while $n = 2$ for black hole
systems (Shakura \& Sunyaev 1973, Fukue 1992, King et al. 1997).

In Fig.~\ref{fig:gam_factor} we show both the net $\gamma$ factor which results
directly from our computations and the $\gamma$ factor from 
Eq.~\ref{eq:gamma2} obtained from the disc thickness distribution in our model.

We see that the irradiation flux computed properly from the model is always
positive, as expected, since the disc is irradiated at least by the nearest 
region. However, in these regions which are not well exposed to the innermost
parts the irradiation  is very small. Effective $\gamma$ factor for 
$\dot m = 0.5$ is only $\sim 2.5 \times 10^{-5}$ in the intermediate disc 
region, between 100 and 10000 \Rs. It becomes larger ($\sim 0.0005 $) 
at the distances $10^4 - 10^5 $ \Rs, as well as in the innermost region,
approaching $\sim 0.04$ at a few \Rs. 

Analytical formula (Eq.~\ref{eq:gamma2}; dashed line in 
Fig.~\ref{fig:gam_factor}) does not provide a good approximation of the 
numerical results. Adopting $n = 2$ in this formula does not lead to better
approximation - in this case the flux is generally underestimated, apart
from the innermost region of a few \Rs.
  
Self-irradiation of optically thick, geometrically thin disc is generally 
inefficient in AGN, as argued already by Tuchman et al. (1990).

\subsubsection{Irradiation due to warm absorber}

An example of the incident radiation flux due to warm absorber is shown in 
Fig.~\ref{fig:flux} (dotted line). Here we assumed the following warm absorber 
parameters:
$R_{min}=1000$, $R_{max}=1200$, 
$\theta_{max}=\pi/3$, $\beta = 0$, $\tau_{tot} =0.6$. 
In this case the irradiating flux
is almost constant up to $ r \approx 1000 $ and falls down as $r^{-3}$ further
out. Therefore, the warm absorber effectively acts as a single point-like 
source which captures and re-emits the fraction of the disc luminosity 
approximately given by $2 \tau  \cos\theta$, and the integration over 
the volume is not essential. 

However, if we assume much larger extension of the warm absorber, the effect
will depend essentially on the density profile.

If the power law index describing the radial density profile, $\beta$, is close
to zero the irradiation flux between $R_{min}$ and $R_{max}$ decreases roughly
as $r^{-1}$, and for $\beta \approx 2$ decreases as $r^{-2}$. Below $R_{min}$
the irradiation flux is constant and above $R_{max}$ decreases as $r^{-3}$.  

Generally, the irradiation by the warm absorber is negligible in the innermost
part of the disc. Self-irradiation is relatively more important but not 
essential, in comparison with the radiation flux dissipated inside the 
accretion disc. At $r \approx R_{min}$, the irradiation by the warm absorber
becomes dominant over self-irradiation and at a few hundred Schwarzschild radii
this irradiation may dominate in the local disc energy budget. The ratio
between the irradiation flux due to the warm absorber and the locally 
dissipated flux roughly saturates at distances larger than the outer radius 
of the warm absorber, and this ratio depends
on the global warm absorber parameters $\tau_{tot}$, $\theta_{max}$, and 
$\beta$. 

Since the effect of self-irradiation becomes again important in the outermost
part of the disc (a few thousands Schwarzschild radii) the relative importance
of the two effects depends on the adopted model parameters.  

\subsection{IR/optical/UV spectra of irradiated accretion discs}

\begin{figure}  
\epsfxsize = 90 mm  
\epsfbox[100 220 350 680]{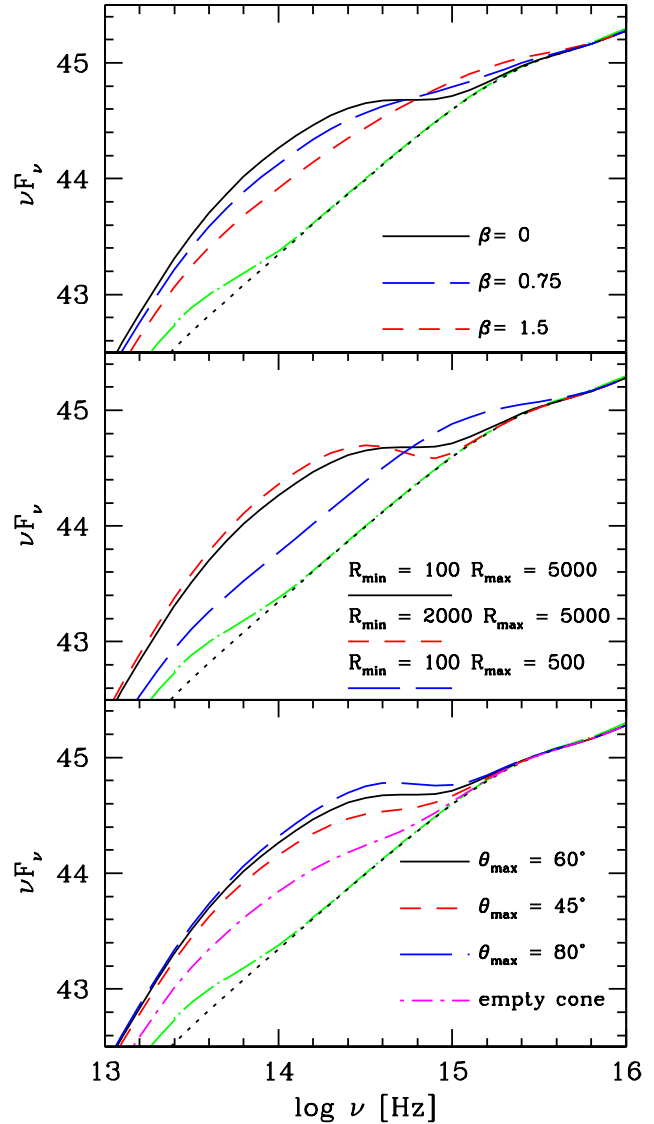}  
\caption{The dependence of the irradiated accretion disc spectrum on the
warm absorber parameters: $\beta$ (upper panel), $R_{min}$ and $R_{max}$
(middle panel) and $\theta$ (lower panel). Parameters of the reference model:
$\dot m = 0.75$, $M = 3 \times 10^7 M_{\odot}$, $\tau_{tot} = 0.6$, $\beta = 0$, 
$R_{min} = 100
R_{Schw}$, $R_{max} = 5000 R_{Schw}$, $\theta_{max} = \pi/3$. Dotted line shows 
the disc spectrum without irradiation, and the long-dash-dot line shows
the effect of the direct irradiation alone.
\label{fig:spectra}}  
\end{figure}  

Examples of the irradiated accretion disc spectra are shown in 
Fig.~\ref{fig:spectra}. The reference model, without irradiation, is marked 
with dotted line. The spectrum is rather hard (blue) in the optical
band. The self-irradiation of the disc changes the spectral slope only in the
IR band, at wavelengths above 3 $\mu$m. 

The indirect irradiation, however, may change the optical spectrum 
considerably.
We see that assuming very slow decrease of the warm
absorber density with distance we obtain spectra which are practically flat
in the optical band. The disc spectrum, however, hardens again in the
 UV band so the effect of irradiation looks like separate spectral component. 
Models with larger $\beta$ still give the spectra 
considerably flatter in the optical range than the standard disc but the 
overall spectrum is practically smooth. 
The spectral range of this flattening depends on the adopted extension of the
warm absorber. 

The amplitude of the effect depends essentially on the warm absorber optical 
depth. If the warm absorber depth $\tau_{tot}$ is of the order of 0.1 
(corresponding to the hydrogen column $N_H \approx 1.5 \times 10^{23}$ 
cm$^{-2}$)  or less, no
effect is visible and the intrinsic disc spectrum is unmodified.

For the model of the warm absorber
adopted in Fig.~\ref{fig:spectra}, the spectral slope in the optical band
$\alpha_o$ is equal to -0.88, -0.56 and -0.20 for $\beta = 0, 0.75$ and 1.5.
Here the convention $F_{\nu} \propto \nu^{\alpha_o}$ is used. 
It compares interestingly with the observed 
slope: Laor et
al. (1997) give the value $\alpha_o = - 0.36$, with dispersion of $0.22$, at 
the basis of the analysis of several PG quasars, and $\alpha_o = - 0.44$ was
given by Vanden Berk et al. (1991) for quasars from SDSS survey.  

The assumed extension of the warm absorber also plays an important role. The
dependence is the strongest if the gas density is uniform (middle panel in
Fig.~\ref{fig:spectra}). If the outer radius is small only the UV part of the
spectrum is modified while for an extended medium the modification of the UV
part is weak while spectra in the optical/IR band are now much flatter.
The affected spectral region is more narrow if $R_{max}/R_{min}$ is close
to 1 and broader if this ratio is large.

The opening angle of the cone occupied by the warm absorber (lower panel
in Fig.~\ref{fig:spectra}) changes mostly the overall normalization of the 
incident flux, similarly to the total optical depth. If this opening angle
is too small ($\theta_{max}< 10^{\circ}$) the presence of the warm absorber 
has no practical effect on the disc spectrum.    

We also considered other types of the warm absorber density distributions.

In one of these special cases we assumed that the density scales with the
radius as $\rho \propto sin^2[(R/R_{min})^{0.6}]$. Such a distribution of
density waves imitates episodes of enhanced and reduced outflow. However,
the resulting spectra were in practice very similar to those obtained from
our power law models for $\beta = 0$. Integration over volume effectively 
smears out any distribution inhomogeneities.

In another of these special cases we assumed the geometry of the warm absorber
in the form of the empty cone centered at the opening angle of 45$^\circ$,
as advocated for quasars by Elvis (2000). In this case the fraction of the
outgoing photons intercepted by the warm absorber is very small 
(model described as 'empty cone' in the lowest panel in Fig.~\ref{fig:spectra}) 
and the change in the spectrum relatively small.

\section{Examples of specific sources}

We choose two interesting sources for detailed modelling. Both are high 
Eddington luminosity objects and belong to the Narrow Line Seyfert 1 class. We 
do not expect an extended inner optically thin flow in this type of objects,
which are analogs of the galactic X-ray sources in their soft state (Pounds et
al. 1995).
Broad band spectra of both sources were extensively studied. One of those, 
RE~J1034+396 (Puchnarewicz et al. 1995, Puchnarewicz et al. 2001), was 
already considered as the best case of 
irradiated accretion disc (Soria \& Puchnarewicz 2002). The other one, 
PG~1211+143, 
is argued to show the presence of the warm absorber of considerable optical 
depth ( Pounds et al. 2003, Gierli\' nski \& Done 2004b). 

\subsection{RE~J1034+396}

\begin{figure}  
\epsfxsize = 90 mm  
\epsfbox{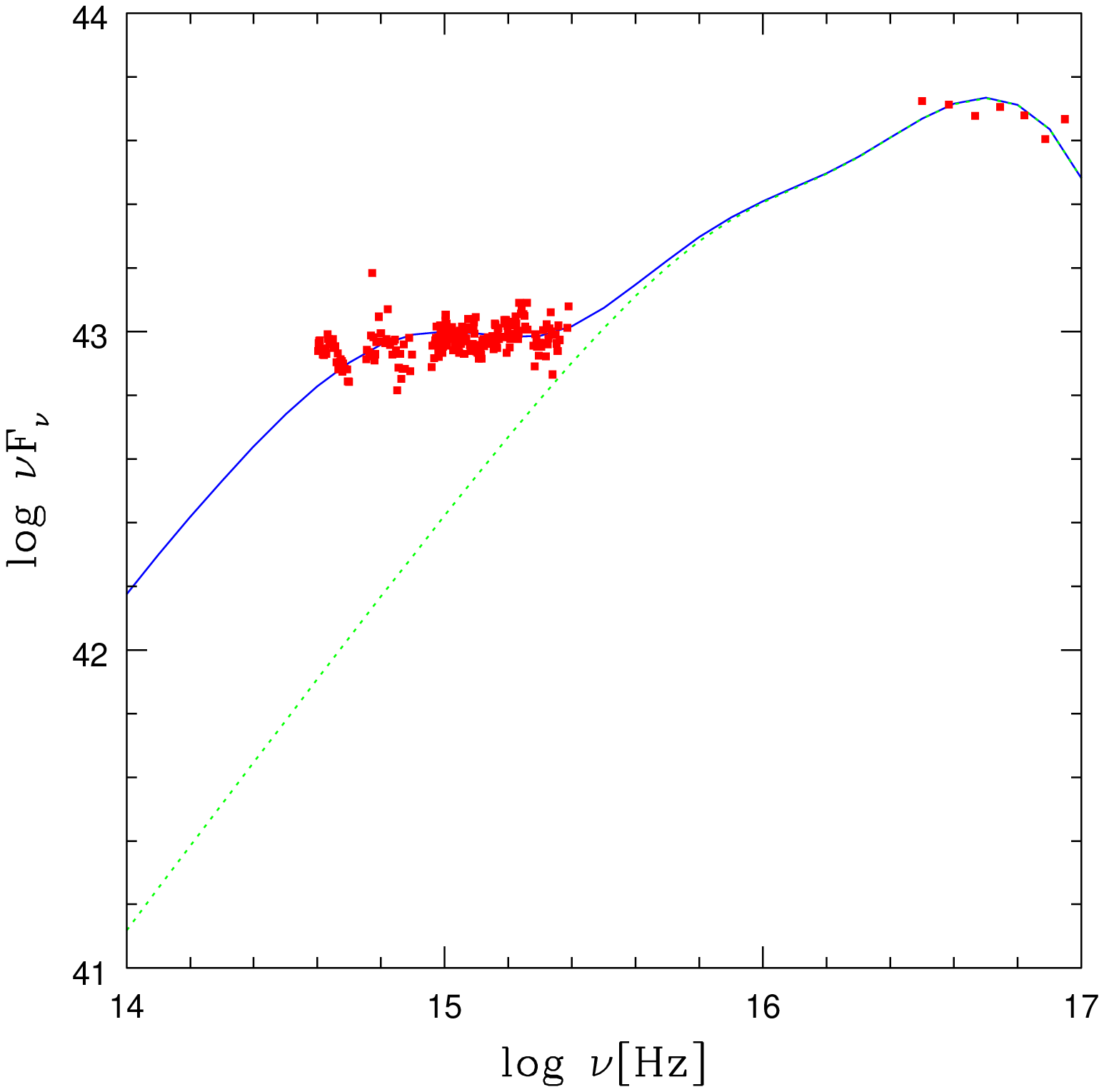}  
\caption{The observational data points for RE~J1034+396 and the
irradiated disc model (continuous line). Parameters of the model:
$\dot m = 0.84$, $M = 6.3 \times 10^5 M_{\odot}$, $\tau_{tot} = 0.6$, 
$\beta = 0.0$, 
$R_{min} = 10
R_{Schw}$, $R_{max} = 7000 R_{Schw}$, $\theta_{max} = \pi/3$.  
Dotted line shows 
the disc spectrum without irradiation. Adopted cosmological model: $H_o=
75$ km s$^{-1}$Mpc$^{-1}$, $q_o = 1/2$.
\label{fig:j1034}}  
\end{figure}  

The observational data for RE~J1034+396 (z=0.04214) was shown by 
Soria \& Puchnarewicz (2002) 
and it was kindly provided
by Liz Puchnarewicz. The nature of the optical/UV component is not 
clear; the spectra do not seem to show the narrow absorption lines typical for
starlight (Puchnarewicz et al. 1995). The source is surprisingly stable in 
comparison to many other NLS1 galaxies (Puchnarewicz et al. 1998) although
hints of variability (in timescales of a fraction of a day) are seen 
in the EUVE lightcurve (Halpern, Leighly \& Marshall 2003).  Warm absorber
features were not discovered so far in this source. However, the source was
observed by Chandra only for 5 ksec, and by XMM only for 15 ksec (according
to http://heasarc.gsfc.nasa.gov/) which is
not enough to detect a highly ionized warm absorber. For example, no warm
absorber was seen in ASCA data and XMM for Ton S180 and even during
the  preliminary analysis of the
Chandra data for the same source but detailed analysis of the 80 ksec 
Chandra data
showed the presence of narrow absorption features (see \Agata et al. 2004). 
Therefore, the
existence of highly ionized warm absorber in  RE~J1034+396 is not excluded.  

Our model of an irradiated accretion disc well represents both optical/UV
continuum and the soft X-ray spectrum (Fig.~\ref{fig:j1034}). 
The mass of the black hole, 
$M= 6.3 \times 10^5 M_{\odot}$ is consistent with the limits 
$0.6 - 3.0 \times 10^6 M_{\odot}$ obtained by Soria \& Puchnarewicz (2002).
The object is accreting still below the Eddington ratio but close to it 
($\dot m = 0.84$). The warm absorber responsible for scattering of the inner
disc emission must be quite extended ($R_{max} = 7000 R_{Schw}$). Such a 
scattering region would smear most of the intrinsic flux variability 
happening in the timescale shorter than a few hours. Longer trends, if present
in the intrinsic flux, should be also seen as corresponding variations in the
optical/UV spectrum.

ASCA and Beppo-SAX data did not show any clear effect of the warm absorber
(Soria \& Puchnarewicz 2002) so the warm absorber in this source, if indeed
present, must be completely ionized. 
However, the presence of strong outflow is rather
expected for sources so close to the Eddington ratio. The soft X-ray spectrum 
of this objects
is well fitted if the disc emission is comptonized by a medium at the 
temperature $T \sim 15 $ keV and optical depth $\tau = 1.3$ under 
assumption of the plane-parallel geometry  
(Soria \& Puchnarewicz 2002). This Comptonizing 
medium may be identified with such outflow, as suggested by Pounds et al. 
(2003)
in the context of PG 1211+143.

Very low value of the parameter $\beta$ of the warm absorber results from the
spectrum being very flat in $\nu F_{\nu}$ plot. If a significant internal 
reddening is present in the source, the actual slope may be positive, and 
$\beta$ larger. Mason et al. (1996) suggested the possible reddening $E_{B-V}
\sim 0.5$ in RE~J1034+396 at the basis of the $H{\alpha}$ to $H{\beta}$ ratio 
of 5, much higher than expected for case B recombination. 

The small blue bump (Balmer continuum+blended iron lines) does not show up
clearly in RE~J1034+396.  

\subsection{PG~1211+143}

\begin{figure}  
\epsfxsize = 90 mm
\epsfbox{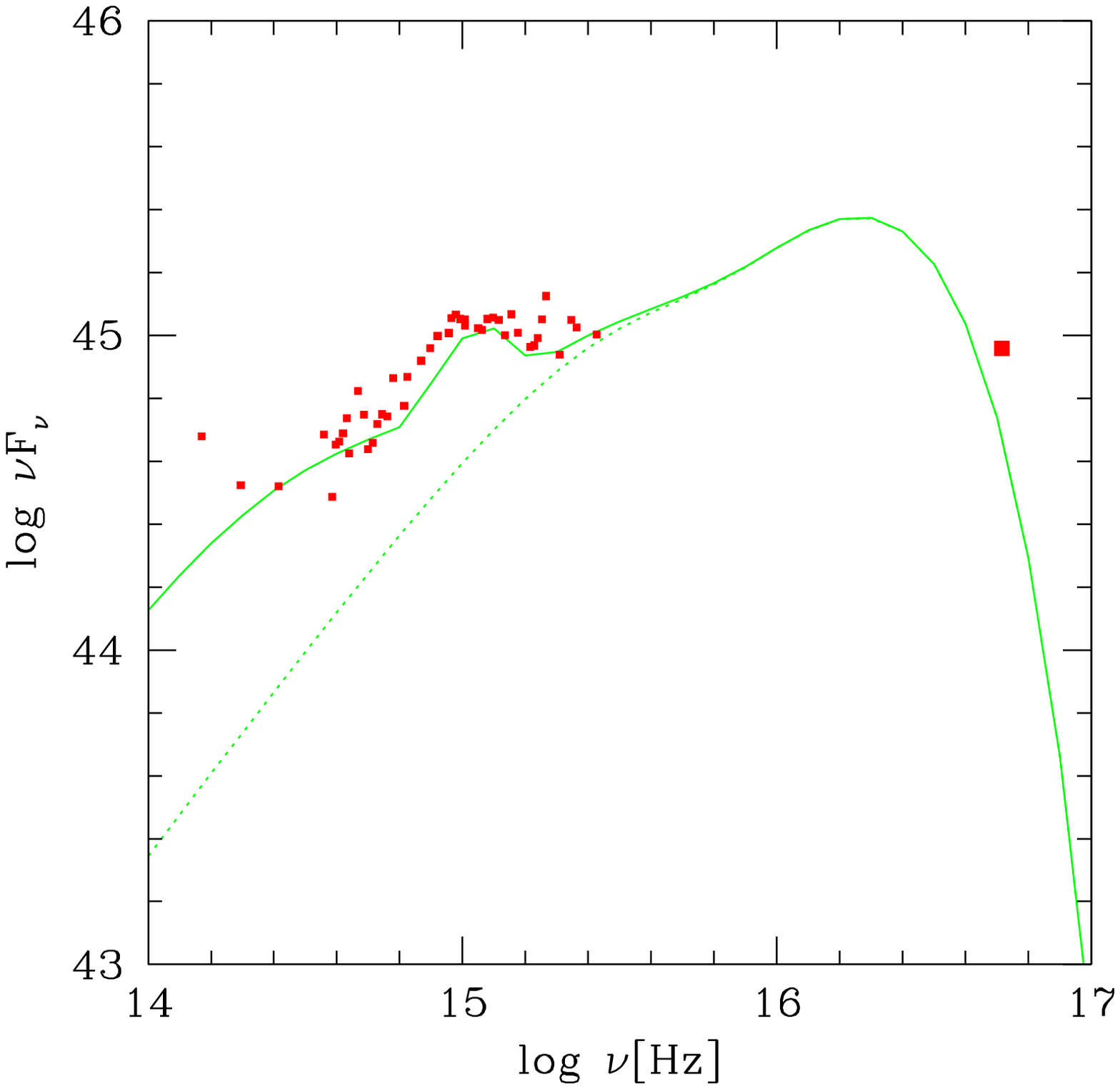}  
\caption{The observational data points for PG~1211+143 and the
irradiated disc model (continuous line). Parameters of the model:
$\dot m = 0.75$, $M = 3.0 \times 10^7 M_{\odot}$, $\tau_{tot} = 0.6$, 
$\beta = 0.75$, 
$R_{min} = 100
R_{Schw}$, $R_{max} = 5000 R_{Schw}$, $\theta_{max} = \pi/3$.  
Dotted line shows 
the disc spectrum without irradiation. Adopted cosmological model: $H_o=
75$ km s$^{-1}$Mpc$^{-1}$, $q_o = 1/2$.
\label{fig:pg1211}}  
\end{figure}  

The observational data for PG~1211+143 (z=0.08090) were taken from 
Elvis et al. (1994).
This source is strongly variable and the presented data are representative
for the periods when the source is bright (see e.g. Janiuk, Czerny \& 
Madejski 2001). 

The shape of the UV continuum in this object clearly indicates the presence of
the small blue bump due to the Balmer continuum and blended iron lines. Such a
component is present in most spectra of Seyfert galaxies. 
Therefore, we modelled
this component and added to the resulting disc spectrum. We adopted the 
spectral shape of the Balmer continuum and the shape of the blended
iron line component after Wills, Netzer \& Wills (1985), 
and we assumed equal amplitude of the two 
components (Neugebauer et al. 1987). 
The overall amplitude of this component was assumed in such 
way that the component constitutes 37 \% of the total
spectrum at 3000 \AA~ and 9 \% at 1890 \AA. The required normalization is 
rather typical for AGN. For example, 
in NGC 5548 the small blue bump accounted for up to 10\% of the flux at 
1840 \AA~ and
30-40\% of the flux at 2670 \AA~ (Peterson et al. 1991, Maoz et al. 1993).  
The result is shown in Fig.~\ref{fig:pg1211}.
 
The mass of the black hole in our model ($M = 3.0 \times 10^7 M_{\odot}$) is 
consistent with the limits obtained by Kaspi et al. (2000) from reverberation
($M_{mean} = 4.05^{+0.96}_{-1.21} \times 10^7 M_{\odot}$, $M_{rms} = 2.36^{+0.56}_{-0.70} \times 10^7 M_{\odot}$), as well as with the results based on the
X-ray variability ($\log M = 7.0 \pm 0.7$ in Janiuk et al. 2001, 
$M = 8.1 \times 10^7 M_{\odot}$ in Niko\l ajuk, Papadakis \& Czerny 2004). The
source is accreting close to the Eddington ratio ($\dot m = 0.75$). The
parameters of the warm absorber are consistent with requirements obtained
by Pounds et al. (2003). Gierli\' nski \& Done (2004b) presented a good 
representation to the broad band data, including X-ray data from XMM-Newton, 
by postulating an outflowing strongly ionized medium with velocity dispersion 
$v/c \sim 0.2$, which Comptonized the disc spectrum. The value of the black 
hole mass used in this analysis ($M = 6.5 \times 10^7 M_{\odot}$, from
Boroson  2002) was somewhat higher than in our model and the Eddington ratio 
correspondingly higher (1.14 from Boroson 2002) since disc irradiation was not 
included in their analysis.

The extension of the warm absorber should greatly reduce any optical 
variability in this source at timescales below $\sim 5$ years.

\subsection{Blue radio quiet quasars}

Large sample of quasars obtained in course of completing Sloan Digital Sky
Survey (hereafter SDSS) allowed to study the optical/UV slope distribution.
The sources were divided into four classes of increasing 'reddening' in their
spectra (Richards et al. 2003) and the composite spectra had the slope 
(measured between 1450 \AA~ 
and 4040 \AA) of -0.25, -0.41, -0.54 and -0.76, correspondingly. This sequence
of spectra was well explained assuming that the effect is due to the increasing
amount of reddening by amorphous carbon dust (Czerny et al. 2004). In this
interpretation no irradiation of the disc seemed to be needed. However,
the spectrum of the composite 1 of Richards et al. (2003), unmodified by dust,
was not fully satisfactorily represented by a bare disc - clear flattening
of the spectrum below $\log \nu \approx 14.9$ was visible.

\begin{figure}  
\epsfxsize = 90 mm  
\epsfbox{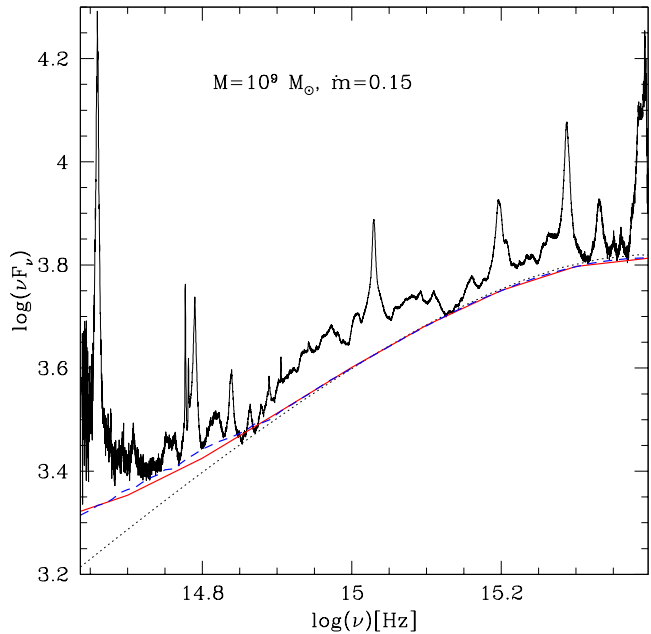}  
\caption{The composite 1 quasar spectrum from SDSS (Richards et al. 2003; 
thin continuous line), disc spectrum without irradiation (dotted line) and disc
illuminated due to the presence of the warm absorber (thick continuous line).
Parameters of the model:
$\dot m = 0.15$, $M = 10^9 M_{\odot}$, $\tau_{max} = 0.6$, 
$\beta = 0.75$, 
$R_{min} = 500
R_{Schw}$, $R_{max} = 700 R_{Schw}$, $\theta_{max} = \pi/3$. Dashed line
shows the non-irradiated disc with starlight.  
\label{fig:SDSS}}  
\end{figure}

We can model this flattening assuming the presence of the warm 
absorber. An example of solution is shown in Fig.~\ref{fig:SDSS}. The model
shows flattening at the longest wavelengths, as requested. However, we did not
find a fully satisfactory representation for this high quality data - the
theoretical spectrum is never quite as flat as requested by the data of 
Richards et al. (2003).

An alternative possibility exists - Vanden Berk et al. (2001)
attributed the change of the 
spectral slope in the quasar composite spectrum at $\sim 5000$ \AA~ 
to starlight contamination, at the basis of the detection
of stellar absorption lines.
Starburst activity may be intrinsically connected with an accretion process.
A number of recent papers discuss the issue of the stars originating/existing
very closely to the active nucleus and interacting directly with the
accretion discs (e.g. Collin \& Zahn 1999, Nayakshin 2004 and the 
references therein). Such a nuclear cluster may contribute to the overall
spectrum. An exemplary spectrum of such a nuclear cluster (see Figure 1. of 
Nayakshin 2004) peaks at $\nu \approx 4 \times 10^{14}$ Hz.

We attempted to model the starlight contribution, taking the spectrum
of the galaxy M31 as a template. Normalization of this contribution was
a free parameter. Such a method was justified by the analysis performed 
by Wamsteker et al. (1990). The result is shown in Fig.~\ref{fig:SDSS}. 

The model with starlight contribution did not
reproduce the quasar spectrum much better than the model
of irradiated disc. It is possible that the use of M31 as 
template is not appropriate (see e.g. Kotilainen \& Ward 1994 for the
discussion of coupling between an AGN and its host galaxy), 
and better model of the circumnuclear cluster
is needed. 

If indeed all flattening at long wavelengths is due to starlight it
means that the outflow of highly ionized material is either inefficient
(low optical depth) or collimated.

\section{Discussion}


The emission from a standard accretion disc was found to be a very natural
model of the optical/UV spectra of AGN (Shields 1978, Czerny \& Elvis 1987,
Wandel \& Petrosian 1988, Sun \& Malkan 1989). However, spectral fits usually
requested the presence of an additional underlying power law component of
unspecified nature since the spectra were flatter than predicted by models,
and this power law did not even seem to be connected to the power law 
component seen in X-ray band (Thompson 1995).

This serious weakness of the disc model caused interests in other models
of the optical/UV bump like optically thin emission (Barvainis 1993).
However, detection of a Balmer edge in a polarized spectrum of a quasar 
Ton 202 by Kishimoto, Antonucci \& Blaes (2003) provided a direct evidence 
in the favour
of the optical/UV emission originating in an optically thick medium (a disc).

However, the discrepancy between the observed optical/UV spectrum and the 
predictions by the canonical models remains. It is not surprising since 
several effects should (in principle) be included in modelling optical/UV
spectra of AGN:
\begin{itemize}
\item {extinction due to circumnuclear dust}
\item {cold disc evaporation in the innermost part and the transition to a hot
flow}
\item {contribution of starlight}
\item {contribution of small blue bump (Balmer continuum and blended iron lines)}
\item {irradiation of the cold disc by the X-ray emission from the 
hot optically thin plasma}
\item {Comptonization of the disc emission by a hot plasma}
\item {irradiation of the disc by scattered disc emission}
\item {self-irradiation of the cold disc}
\item {proper local models of irradiated disc atmosphere}
\item {non-stationarity of accretion flow and of X-ray generation.} 
\end{itemize} 
Complexity of the problem is depressing, and in many cases it is difficult to
disentangle various effects. However, some help comes from analyzing broad band
spectra since information about the X-ray emission provides constraints to
factors modifying optical/UV band.

In the present paper we model
the scattering of the disc radiation by an extended ionized medium and 
redirecting its emission back again towards the disc. The observational 
evidence of such a medium comes from the studies of warm absorbers 
in Seyfert galaxies (e.g. Ashton et al. 2004 and the references therein) 
and Broad Absorption Line (BAL) 
quasars (e.g. Gallagher et al. 2004 and the references therein).
Observations of these objects give direct measurements of the column 
density of the partially ionized material, being less sensitive to the fully
ionized plasma. Polarimetry observations of Seyfert 2 galaxies with hidden
Broad Line Regions (Antonucci \& Miller 1985; Tran 2001) 
and of BAL QSO (e.g. Lammy \& Hutsemekers 2004) 
give in turn direct information about the
'scatterers' but at unspecified (rather large) distances from the central 
black hole. Estimates of the total column density of the material along the 
line of sight obtained from the UV studies are frequently inconsistent, and the
amount of fully ionized material can be frequently underestimated.
The detection rate of the warm absorber decreases with an increase
of the object accretion rate (e.g. Laor et al. 1997), BAL features are seen
in bright QSO. Although the connection between BAL outflow and warm absorbers
is not clear they may represent basically the same outflow but of different
ionization state and/or seen at 
different inclination angles.

This material can be traced, however, by modelling the optical/UV spectra.
We show that if the warm absorber has considerable optical depth 
$\tau_{tot} > 0.1$ (or $N_H > 10^{23}$ cm$^{-2}$), and if this medium
is located relatively close to the central region we can expect significant 
change in
the optical/UV spectrum of the accretion disc in such source due to
the scattering of the fraction of radiation by this warm absorber and
redirecting a fraction of this emission back towards the disc. The
effect depends not only on the total optical depth of the warm absorber
along the line of sight but on the whole density distribution.

We considered in detail the case of a disc irradiated due to the
presence of the warm absorber but we neglected disc evaporation and
irradiation by hard X-ray source. We showed that the direct 
self-irradiation of the disc  in this case is not very efficient 
but the indirect irradiation due to the scattering by the warm absorber may
flatten the optical spectrum considerably. 

In our opinion the model directly applies to high accretion rate AGN, 
$\dot m > 0.2 $ 
like quasars or Narrow Line 
Seyfert 1 galaxies. In those sources the disc extends practically down to
the marginally stable orbit (Pounds et al. 1995, 
Sobolewska, Siemiginowska \& \. Zycki 2004), and 
an extended ADAF does not form.
The hard X-ray source contains a small fraction of the source bolometric 
luminosity, as the optical/UV/soft X-ray Big Blue Bump strongly dominates the
overall spectrum. Therefore, the irradiation by hard X-rays is 
energetically inefficient although it is important from the point of view of
the reflection component formation. Additionally, if the hard X-ray emission
comes from coronal flares and it is confined to the direct vicinity of the
disk, our assumptions of emission comming from the disk surface are roughly 
satisfied even for this component.
  
We obtained satisfactory representation of the data for two NLS1 galaxies,
RE J1034+396 and PG1211+143. Blue quasar composite 1 spectrum of Richards
et al. (2003) is almost well represented by a standard disc, with some
flattening only above $\lambda \sim 5000 $ \AA. We modelled this flattening
also as the effect of irradiation due to warm absorber but the result
was not quite convincing and the contribution of starlight seems to be
most probable explanation.

We neglected other possible effects from our
list although intrinsic reddening as well as Comptonization 
can strongly bias the observed FUV spectral index (Shang et al. 2004).

The irradiation of the disc due to the presence of the warm absorber may
also be important in Seyfert 1 galaxies.
However, typical Seyfert 1 galaxies may have 
their discs truncated at
larger radius, and a geometrically thick inner flow develops (e.g. 
Poutanen, Krolik, \& Ryde 1997,
Chiang \& Blaes 2003), the
bolometric luminosity of the hard X-ray component is larger, and  the 
source is most probably located at large distance from the equatorial plane, 
either corresponding to the disk truncation radius (in ADAF models), 
or to the position of a base of
jet (Henri \& Pelletier 1991; Martocchia, Matt \& Karas 2002). 
It enhances the direct self-irradiation by the
central X-ray emitting region (Chiang \& Blaes 2003). 
So it is possible that in Seyfert galaxies both direct X-ray irradiation 
and indirect irradiation due to the warm absorber scattering 
can be equally important, but the model would include additional
parameters specifying the location of the X-ray emitting plasma.

\section*{Acknowledgements}  
 
This work was  
supported in part by grant 2P03D~003~22 (BCz \& ZL).
of the Polish State  
Committee for Scientific Research (KBN).
The project made use of the code {\sc cloudy}, version 94.0 of Gary Ferland 
(http://www.nublado.org/). This research has made use  
the High Energy Astrophysics Science Archive Research Center Online 
Service, provided by the NASA/Goddard Space Flight Center.

\ \\  
This paper has been processed by the authors using the Blackwell  
Scientific Publications \LaTeX\  style file.  
 
\end{document}